\documentclass{ws-procs9x6}

\begin{document}

\title{Proceedings for TASI 2009 Summer School on \\
``Physics of the Large and the Small'' :\\
Introduction to the LHC experiments}

\author{E. HALKIADAKIS$^*$}

\address{Department of Physics \& Astronomy, Rutgers University,\\
Piscataway, New Jersey 08854, U.S.A.\\
$^*$E-mail: evahal@physics.rutgers.edu\\
www.physics.rutgers.edu}

\begin{abstract}
These proceedings are a summary of four lectures given at the
Theoretical Advanced Study Institute in Elementary Particle Physics
(TASI) in 2009.  These lectures provide a basic introduction to
experimental particle physics and the Large Hadron Collider
experiments at CERN, with many general examples from the (still
running) Fermilab Tevatron.
\end{abstract}

\keywords{Elementary Particle Physics; Collider Physics}

\bodymatter

\section{Introduction}\label{sec:intro}

\begin{figure}[t]
\begin{center}
\psfig{file=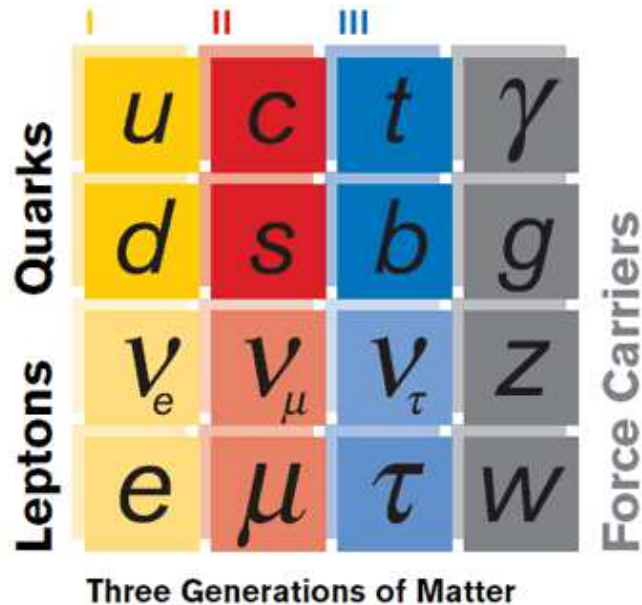,width=3.5in}
\caption{The Standard Model. \cite{sm}}
\label{figs:SM}
\end{center}
\end{figure}

The Standard Model (SM) of elementary particles, summarized in
Fig.~\ref{figs:SM}, has been quite successful in making predictions,
confirmed to incredible precision in experimental data.  Yet there are
still many unanswered questions about nature and the fundamental
interactions.  Some of today's most challenging questions in physics
are, but not restricted to:

\begin{itemize}
\item Is there really a Higgs boson, as predicted by the Standard Model of particle physics?  If so, what is its mass?  
\item If not, what is the origin of electroweak symmetry breaking?
\item Why is there a hierarchy of masses?
\item What are the origins of dark matter and dark energy?
\item Why is there no anti-matter in the universe?
\item How does gravity fit into all this?
\end{itemize}

The dawn of a new energy frontier has arrived with the
recent turn-on of CERN's Large Hadron Collider (LHC). The LHC
experiments at CERN use state-of-the-art technology and will hunt for
answers to many of the open questions in high energy particle physics
today.  From the discovery potential of the Higgs boson, to new
particle and new phenomena searches, all particle physicists are
focused on upcoming LHC results.

What follows is a summary of four lectures given at TASI in
2009~\cite{tasi09}.  These lectures provide a basic introduction to
experimental particle physics, with an emphasis on CERN's LHC
experiments.  First, I begin with an overview of particle accelerators
(Sec.~\ref{sec:accel}) with an emphasis on the currently running
hadron colliders, the Fermilab Tevatron and the LHC.  Next, I review
the importance of luminosity (Sec.~\ref{sec:lumi}), the proton
composition (Sec.~\ref{sec:proton}) and hadron
collisions (Sec.~\ref{sec:hadroncoll}), followed by a summary of a few
key definitions every high energy physics should know
(Sec.~\ref{sec:defs}). I then review how particles interact with
matter (Sec.~\ref{sec:partint}) and how those interactions are used in
designing particle detectors (Sec.~\ref{sec:partdet}) and the
identification of particles for analysis (Sec.~\ref{sec:partid}).
Finally, I describe the importance of a trigger (Sec.~\ref{sec:trig}),
the current status of the LHC (Sec.~\ref{sec:lhcstat}), I highlight a
few of the early LHC physics measurements expected
(Sec.~\ref{sec:lhcphys}) and conclude (Sec.~\ref{sec:conc}).

\section{Particle Accelerators}\label{sec:accel}

Particle accelerators are shaped in one of two ways: 
\begin{itemize}
\item {\it Linear colliders or LINAC:}  An example of such an accelerator is
 the Stanford Linear Accelerator Center (SLAC). 
\item {\it Circular or synchrotron accelerators:} These provide higher
  energies than a LINAC, such as the Fermilab Tevatron.
\end{itemize} 

\noindent Accelerators can also be arranged to provide collisions of two types: 
\begin{itemize}
\item {\it Fixed target experiments:} When particles are shot at a fixed target. 
The center-of-mass energy, $\sqrt{s}$, for this class of experiments is:
\begin{eqnarray}
\sqrt{s} = \sqrt{2 ~ E_{beam} ~ m_{target}} 
\end{eqnarray}

\item {\it Colliding beam experiments:} When two beams of particles are 
made to cross each other.  In this case, 
\begin{eqnarray}
\sqrt{s} = 2 ~ E_{beam} 
\end{eqnarray}
\end{itemize}

\noindent Circular accelerators have been arranged to collide
electrons and positrons (for example at LEP) and protons and
(anti-)protons (hadron colliders). Scattering experiments have been
also done by colliding leptons (electrons or positrons) and protons
(for example at HERA).  Examples of hadron colliders are the Tevatron
at Fermilab or the Large Hadron Collider at CERN.  Hadron colliders
provide much higher energies than $e^+e^-$ colliders and do not suffer
from synchrotron radiation.  However, $e^+e^-$ colliders can provide
us with clean environment for precision measurements.

The two currently running hadron colliders, the Tevatron and the LHC,
are further described in the following sections.

\subsection{Fermilab Tevatron}

The Fermilab Tevatron, located roughly 30 miles west of Chicago, IL,
accelerates protons and anti-protons to $\sqrt{s}=1.96$ TeV.  The main
ring is roughly 4 miles in circumference and when running collides 36
bunches of protons against 36 bunches of anti-protons, with roughly
100 billion particles in each bunch.  Once injected, the beam is
stored and the same bunches are collided typically for 20-30 hours.

The Tevatron hosts two ``general purpose'' experiments, the Collider
Detector at Fermilab (CDF) and DO.  Run I of the Tevatron
lasted from 1992-1996 and in 1995 the two experiments announced the
discovery of the top quark.  At that point, the Tevatron entered a
fixed target phase and then 2001 marked the start of Run II and will
continue until at least FY2011.

\subsection{Large Hadron Collider}

The Large Hadron Collider at the CERN laboratory near Geneva, Switzerland,
is a proton-proton collider with a 27 km circumference.  It is
designed to provide collisions with a maximum $\sqrt{s}=14$ TeV.  In
November-December 2009, the LHC turned on and collided protons at
$\sqrt{s}=900$ GeV and for the first time at $\sqrt{s}=2.36$ TeV,
exceeding the center-of-mass energy of the Tevatron.  On March 30,
2010 the LHC achieved collisions at $\sqrt{s}=7$ TeV, launching a new
era in particle physics.  The LHC will also collide heavy ions (Pb-Pb)
for shorter running periods of roughly 1 month per year.

\begin{figure}[t]
\begin{center}
\psfig{file=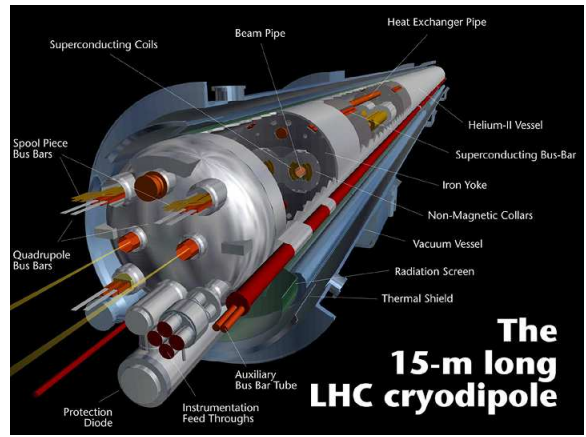,width=3.in}
\caption{Computer generated diagram of an LHC dipole magnet.~\cite{lhc}}
\label{figs:dipole}
\end{center}
\end{figure}

The LHC tunnel rests 100 meters underground.  The beams circle the
ring inside vacuum pipes guided by super-conducting magnets.  There are
thousands of magnets directing the beams around the accelerator,
including 1232 15 meter long, 35 ton dipole magnets shown in
Fig.~\ref{figs:dipole}.  These dipole magnets have an ingenious
configuration called a ``2-in-1'' design allowing the two proton beams
to point in opposite directions in each pipe.  For a 7 TeV energy
beam, the dipoles are cooled to a temperature of 1.9$^{o}$ K providing
an 8.4 T magnetic field and a current flow of 11.7kA.

The LHC is designed to collide a maximum of 2808 proton bunches
against another 2808 proton bunches.  Each bunch is several cm long
and contains approximately 100 billion protons.  In order to increase
the probability of a hard collision, the beam is squeezed as much as
possible at the interaction point to a diameter of tens of microns.
For these operating design conditions, it is expected that on average
20 additional $pp$ interactions will occur.  

\begin{figure}[t]
\begin{center}
\psfig{file=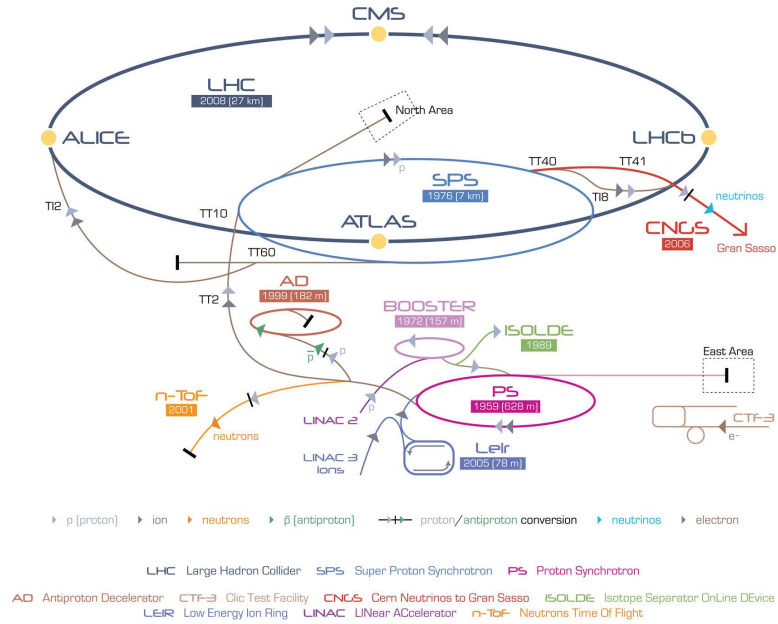,width=4.in}
\caption{The LHC accelerator chain.~\cite{lhc}}
\label{figs:lhcacc}
\end{center}
\end{figure}

The LHC accelerator chain is shown in Fig.~\ref{figs:lhcacc}.
Initially, 50 MeV protons are produced in the LINAC and accelerated to
1.4 GeV in the Booster.  They are then injected in the Proton
Synchrotron (PS) where they reach an energy of 26 GeV and are further
accelerated to 450 GeV in the Super Proton Synchrotron (SPS).
Finally, they are injected in the main ring where they reach a maximum
energy of 7 TeV (the maximum to-date has been 3.5 TeV per beam).

The collisions at the LHC take place at the location of the four experiments, which are:
\begin{itemize}
\item{Compact Muon Solenoid (CMS): One of the two large ``general purpose'' experiments.}
\item{A Toroidal LHC Apparatus (ATLAS): The other of the two large ``general purpose'' experiments.}
\item{LHCb: Designed to study the $b$-quark sector, CP violation and rare decays.}
\item{A Large Ion Collider Experiment (ALICE): A heavy ion experiment designed to study the nature of quark-gluon plasma.}
\end{itemize}

These lectures focus primarily on the CMS and ATLAS
detectors. Finally, Tab.~\ref{tab:lhctev} shows a summary of the LHC
and Tevatron parameters for comparison.

\begin{table}[t]
\tbl{Comparison of the LHC and Tevatron accelerator statistics.}
{\begin{tabular}{c c c} \hline
 & LHC (design) & Tevatron (achieved) \\ \hline
Center-of-mass energy & 14 TeV & 1.96 TeV \\
Number of bunches & 2808 & 36 \\
Bunch spacing & 25ns & 396ns \\
Energy stored in beam & 360MJ & 1MJ \\
Peak Luminosity & $10^{33}-10^{34}cm^{-2}s^{-1}$ & $3.87 \times 10^{32}$ (April 2010)  \\
Integrated Luminosity / year & 10-100 $fb^{-1}$ &  $> 2 fb^{-1}$ (2008) \\ \hline
\end{tabular}
}
\label{tab:lhctev}
\end{table}

\section{Luminosity} \label{sec:lumi}

Important parameters in colliders are the energy of the beams and the
rate of collisions ($R$), or the luminosity ($\mathcal{L}$).  $R$, is defined as:

\begin{eqnarray}
R &=& {\frac{dN}{dt}} = \mathcal{L} \sigma, 
\end{eqnarray}

\noindent where $\frac{dN}{dt}$ is the number of hard collision
events produced per second, and $\sigma$ is the cross section of the process produced.  Integrating over time, we get:

\begin{eqnarray}
N_{\rm{events ~~ produced}} &=& \sigma \times \int \mathcal{L} dt, 
\end{eqnarray}

\noindent where $N_{\rm{events ~~ produced}}$ are the number of
produced hard collision events of the process with cross section
$\sigma$ and $\int \mathcal{L} dt$ is the integrated luminosity which
is provided by the accelerator in a given time period.  Unfortunately,
a given high energy physics detector does not observe every collision
event that is produced.  For example, the trigger is inefficient, as
is the identification of the final state particles, and some fraction
of the events may be produced beyond the detector acceptance (see
Sec.~\ref{sec:xsec}).  These inefficiencies need to be experimentally
evaluated and once accounted for the expression becomes:

\begin{eqnarray}
N_{\rm{events ~~ observed}} &=& \sigma \times \int \mathcal{L} dt \times \epsilon 
\end{eqnarray}

\noindent where $N_{\rm{events ~~ observed}}$ is now the number of
events observed in the detector, and $\epsilon$ is the total
efficiency of identifying the collision event of interest (see
Sec.~\ref{sec:xsec}).

The units of a cross section are the same as the units of area and in
high energy physics are typically represented by a barn (1 barn =
$10^{-24} cm^2$), for example, $mb, \mu b, nb$, etc.  The units of
{\it instantaneous} luminosity are the same as the units of [1 /
  (cross section $\times$ time)], for example $cm^{-2} s^{-1}$.  {\it
  Integrated} luminosity has units of [1 / cross section], for example
$cm^{-2}$ or $pb^{-1}$, $fb^{-1}$, etc.

\begin{figure}[t]
\begin{center}
\psfig{file=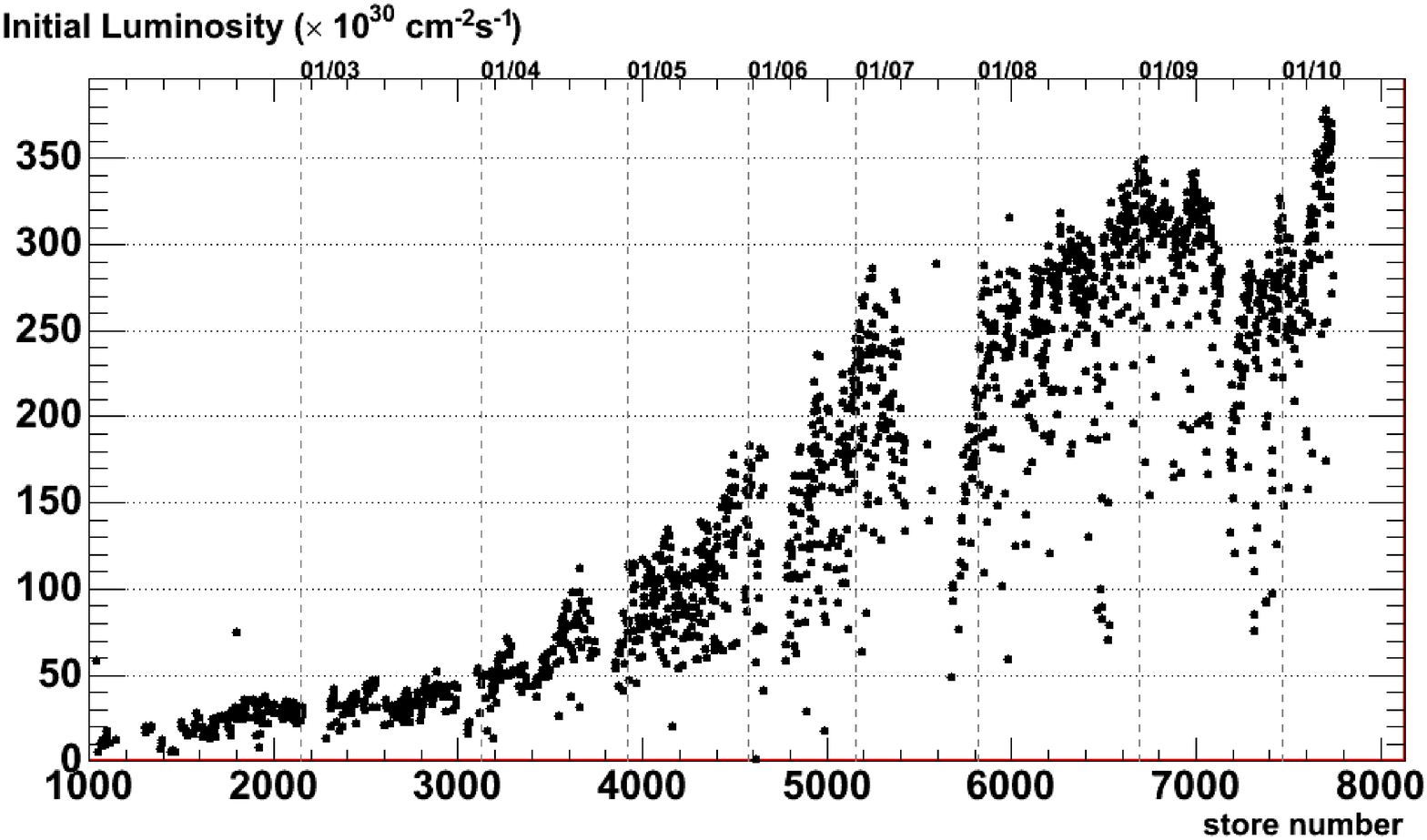,width=4.in}
\psfig{file=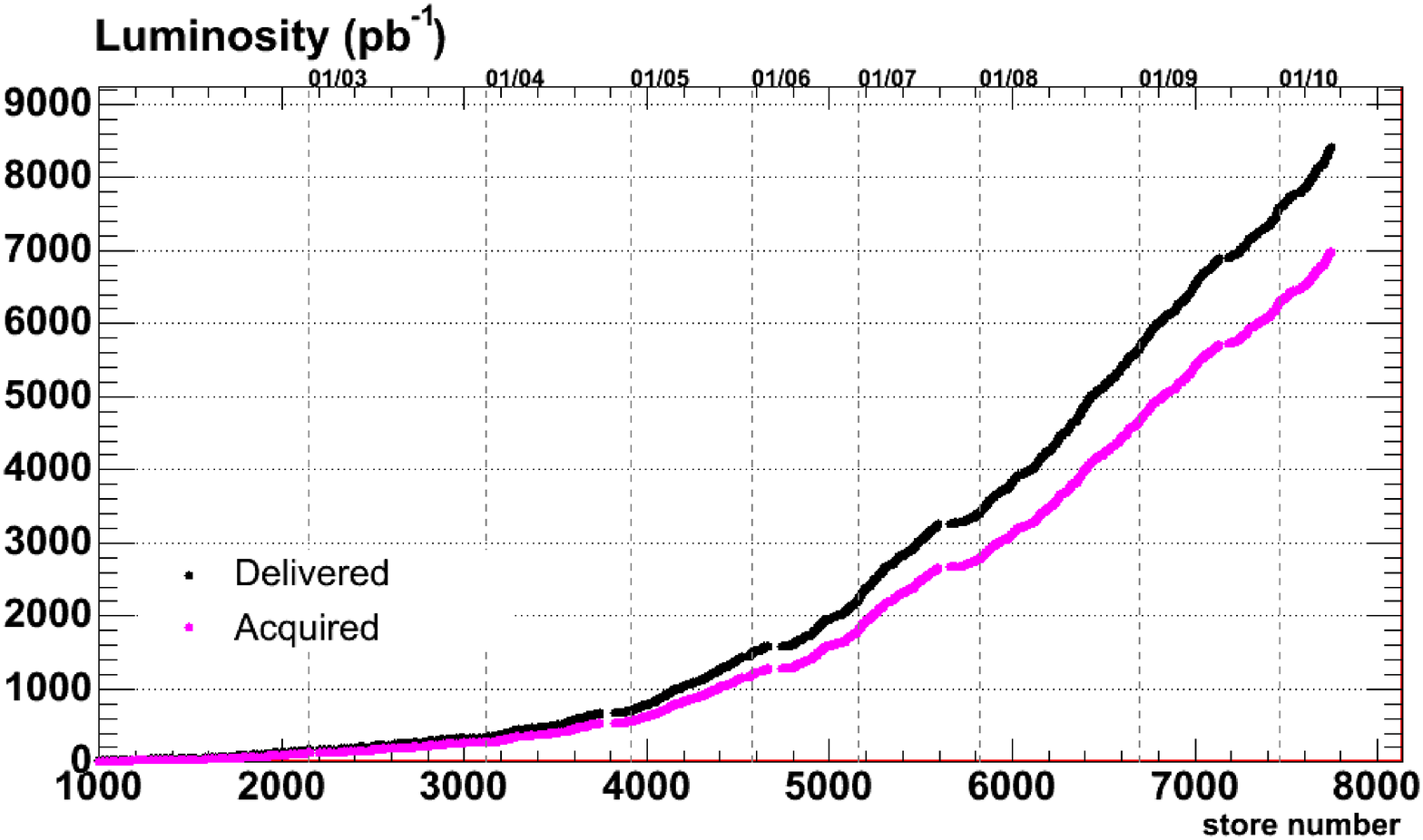,width=4.in}
\caption{Top: Initial luminosity in $(\times 10^{30} cm^{-2} s^{-1})$
  delivered by the Tevatron vs. time. Bottom: Integrated luminosity in
  $pb^{-1}$ delivered by the Tevatron (black) and acquired by the CDF
  experiment (purple) vs. time. ~\cite{cdf}}
\label{figs:lumi}
\end{center}
\end{figure}

An example of the difference between integrated and instantaneous
luminosity is shown in Fig.~\ref{figs:lumi}.  The top figure shows the
initial luminosity delivered by the Tevatron versus time and the
increasing slope demonstrates the challenges of increasing the
luminosity at a hadron collider.  It should be noted that the
instantaneous luminosity drops as the protons collide, until the next
store or fill is dropped followed by (anti-)protons being re-injected
and collisions resume.  The bottom of Fig.~\ref{figs:lumi} shows the
integrated luminosity delivered by the Tevatron (black) and that
acquired by the CDF experiment (purple) as a function of time; it is
impossible to record every collision at a hadron collider and the
difference between the two curves shows how efficiently the experiment
(in this case CDF) collects the data that the accelerator delivers.

Next, let us consider an alternate expression for luminosity:
\begin{eqnarray}
\mathcal{L} = f \frac{n_1 n_2}{4 \pi \sigma_x \sigma_y} \approx f \frac{n_b N_p^2}{4 \pi \sigma_x \sigma_y},
\label{eqn:lumi}
\end{eqnarray}

\noindent where $n_1$ and $n_2$ are the number of particles (protons)
in each of the colliding bunches, $f$ is the frequency with which they
collide, $\sigma_x$ and $\sigma_y$ represent the size of the
transverse beam (e.g. the RMS if we assume a Gaussian shaped beam),
$n_b$ is the number of bunches and $N_p$ is the number of particles
per bunch.  So in order to increase the luminosity, it is important to
{\it squeeze} as many protons in as small a transverse beam spot as
possible.

\subsection{Exercises}

\begin{enumerate}
\item Imagine a hadron collider such as the LHC or the Tevatron runs
  for one year with and {\it instantaneous} luminosity of $10^{31}
  cm^{-2} s^{-1}$, how much {\it integrated} luminosity will be
  delivered to an experiment?  \\
  {\it Answer:}  A year is $ 3 \times 10^7$ seconds, however, accelerators do not operate every day.  
  Assuming a good year of running is $10^7$ seconds, we get a rough estimate:
  \begin{eqnarray}
  \int \mathcal{L} dt = 10^{31} cm^{-2} s^{-1} \times 10^7 s = 10^{38} cm^{-2} = 10^{14} barns = 100 pb^{-1} \nonumber
  \end{eqnarray}

\item In 100 $pb-1$ of data, how many $p \bar{p} \rightarrow t \bar{t}$
  events will be produced at the LHC at $\sqrt{s}= 7$ TeV?  \\
  {\it Answer:} The $p \bar{p} \rightarrow t \bar{t}$ cross section at 7 TeV is $\sim 165 pb$.~~\cite{kidonakis}
  \begin{eqnarray}
  N_{\rm{events ~~ produced}} &=& \sigma \times \int \mathcal{L} dt  \nonumber \\
  &=& 165 pb \times 100 pb^{-1} =  16,500 ~~ t \bar{t} ~~{\rm pairs} \nonumber
  \end{eqnarray}
  Precisely how many events are observed depends on the efficiency of
  observing them in the detector.

\item What size beam spot is needed for $\mathcal{L} = 1 \times
  10^{34} cm^{-2} s^{-1}$ at the LHC? \\ {\it Answer:} The LHC machine
  frequency is $f = c/27$ km = 11kHz, and is designed to contain $n_b
  = 2808$ bunches and $N_p = 1 \times 10^{11}$ protons per bunch.
  Substituting this into Eq.~\ref{eqn:lumi} above and solving for
  $\sigma$ (assuming $\sigma_x \approx \sigma_y$) gives:
  \begin{eqnarray}
   \sigma_{x,y} = \sqrt{11 kHz \frac{(2808) (10^{11})^2}{4 \pi (10^{34} cm^{-2} s^{-1}) } } = 1.5 \times 10^{-3} cm = 15 \mu m \nonumber
  \end{eqnarray}
  So we will need approximately 15 $\mu m$ beam size. For comparison,
  the Tevatron beam size is $\sim 35 \mu m$.

\end{enumerate}

\section{Proton Composition} \label{sec:proton}
\begin{figure}[t]
\begin{center}
\psfig{file=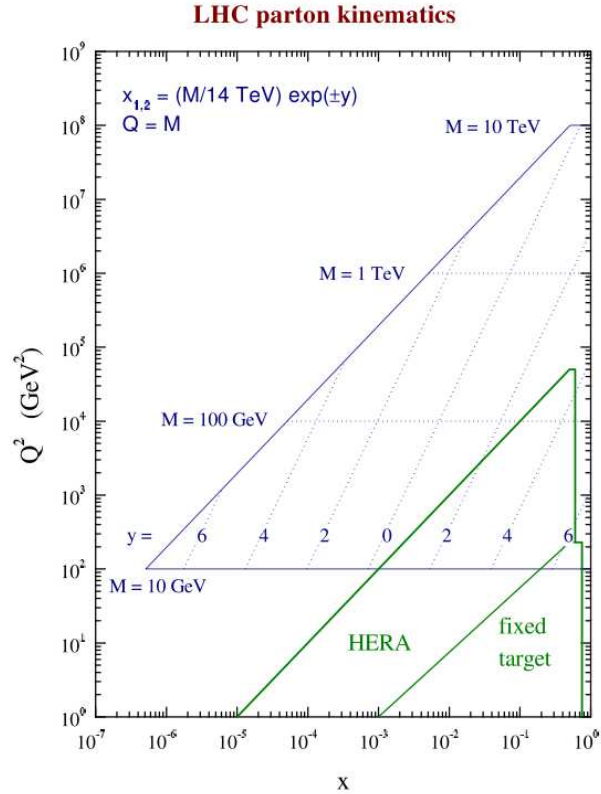,width=3.5in}
\caption{PDF constraints from global fits to data shown as a function
  of parton variables $Q^2$ vs. $x$ (green).  Also shown is the
  relationship between these parton variables and the kinematic
  variables for a final state produced with mass $M$ and rapidity $y$
  assuming and LHC energy $\sqrt{s}=14$ TeV (blue). ~\cite{pdf}}
\label{figs:parton}
\end{center}
\end{figure}

The proton is composed of three valence quarks (two up quarks and one
down quark) as well as gluons and sea quarks, but the exact
composition is quite complicated.  The mixture of partons inside the
proton depends on the Bjorken-$x$ (the fraction of the proton's
momentum carried by the parton) and $Q^2$ (the momentum scale that
characterizes the hard scattering, such as $M^2$, where $M$ is
the mass of the particle that is created by the scattering process).
These quantities, $x$ and $Q^2$, are also what parameterize Parton
Distribution Functions (PDF's), as seen in
Fig.~\ref{figs:parton}~\cite{pdf}, which help describe the content
of the proton.  For low values of $Q^2$ ($Q^2 <$ 1GeV$^2$) the proton
behaves predominantly as a single particle.  For a medium energy range
(1 $< Q^2 < 10^4$ GeV$^2$), the proton interacts as a composite
particle and the valence quarks dominate in the interaction. At higher
energies, the gluons and sea quark PDF's are dominant.  PDF's are
obtained by global fits to data measurements from many experiments
(deep inelastic scattering, fixed target, collider) and the
constraints are summarized in Fig.~\ref{figs:parton} (green).  They
are essential inputs to perturbative calculations of production cross
sections at hadron colliders.  There are two main PDF fitting groups,
CTEQ~\cite{cteq6m} and MRST (now MSTW)~\cite{mrst2006,mstw}, which
regularly provide updates to the PDF fits and their uncertainties with new data.

\begin{figure}[t]
\begin{center}
\psfig{file=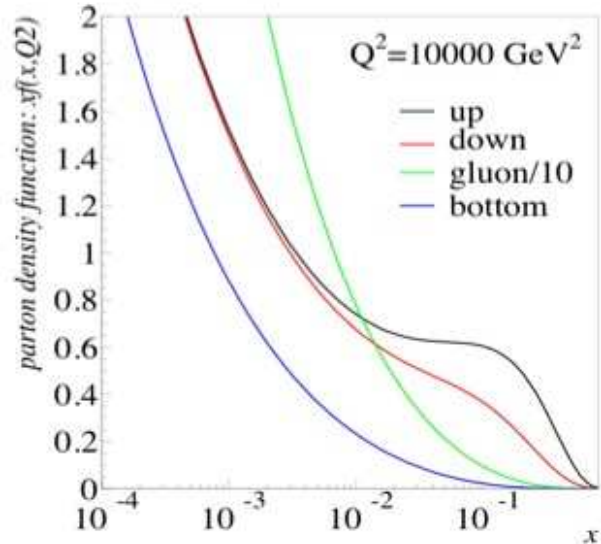,width=3.5in}
\caption{PDF vs. $x$ for up and down valence quarks, bottom sea quarks,
  and gluons (times 0.1) for a $Q^2 = 10000$ GeV$^2$.  The PDF's shown
  are CTEQ6.1M ~\cite{cteq6m} and taken from this ~\cite{pdfmaker} useful
  website. }
\label{figs:pdf}
\end{center}
\end{figure}

Figure~\ref{figs:pdf} shows the PDF's vs. $x$ for the valence quarks
(up and down), sea quarks (upbar) and gluons (divided by a factor of
10).  Note that the PDF's have a dramatic rise at low values of $x$ and
are dominated by gluons in that region.  The valence quarks are
dominant for roughly $x >0.1$.  Uncertainties in PDF's quantify our
understanding of parton content of protons and the cross sections of
processes.  Therefore, making measurements which are sensitive to
constraining PDF's are important since large uncertainties in PDF's
result in large uncertainties in predictions and processes which are
not well understood.  PDF uncertainties can vary quite a lot (roughly
2-30\% or more) depending on the $x$ range and parton of
interest.~\cite{cteq6m,mrst2006,mstw} For example, gluon PDF's are
poorly constrained in the range approximately $x >0.1$.

\section{Hadron Collisions} \label{sec:hadroncoll}
\begin{figure}[t]
\begin{center}
\psfig{file=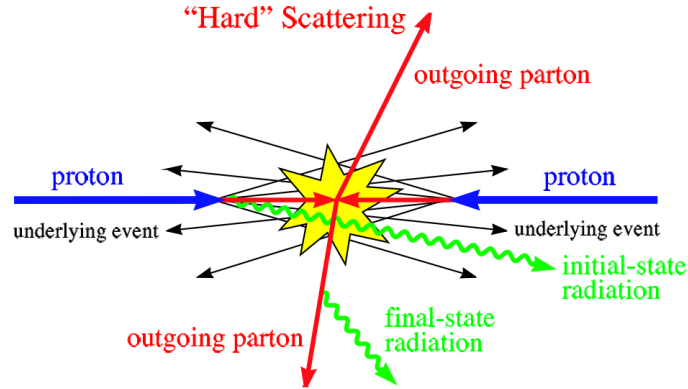,width=4.in}
\caption{Schematic of a hard scattering proton-proton collision.~\cite{pdf}}
\label{figs:ue}
\end{center}
\end{figure}

The collisions, or scattering, which occurs in hadron colliders is
separated into hard and soft scattering.  Calculations of the hard
scattering process (when two of the constituent partons in the proton
collide head-on) are done using perturbative QCD.  The soft processes
(elastic, single diffractive, double diffractive and non-diffractive
inelastic scattering) are much more difficult to understand and suffer
from non-perturbative QCD effects.  The majority of the total $pp$
collisions are soft.  These soft processes (everything except the hard
scatter) is also generally referred to as the ``underlying event''.
The underlying event includes initial state radiation, final state
radiation and interactions of other remnant partons in the proton.  A
schematic diagram describing the hard and soft processes in a hadron
collision can be seen in Fig.~\ref{figs:ue}.  Additionally, there is a
lot about the collision which we do not know, such as which partons
collided with each other, what the momentum of the partons were when
they collided, and what was the effect of the other partons in the
proton.

\begin{figure}[t]
\begin{center}
\psfig{file=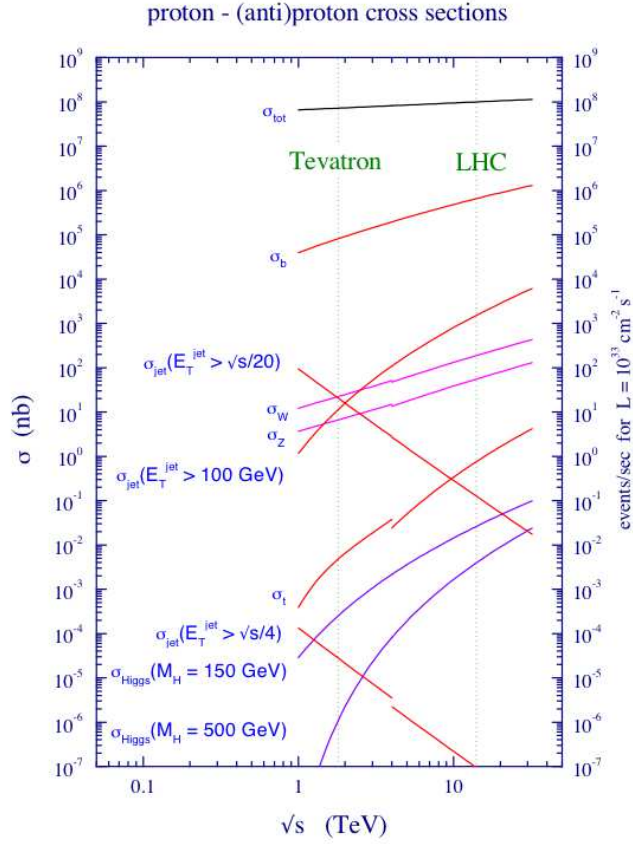,width=3.5in}
\caption{Cross sections (left y-axis) and event rates (right y-axis)
  for SM processes for proton-(anti-)proton collisions as a function
  of center-of-mass energy.~\cite{pdf}}
\label{figs:smxsec}
\end{center}
\end{figure}

Figure~\ref{figs:smxsec} shows the cross sections for various SM
processes as a function of $\sqrt{s}$.  The two vertical lines at $\sim$
2 TeV and 14 TeV represent the Tevatron and LHC energies,
respectively, and note a dramatic increase in many of the cross
sections for the increased $\sqrt{s}$.  This also again emphasizes that
the majority of the total inelastic cross section is coming from soft
scattering processes rather than hard collisions. For example, reading
from the left side of the y-axis, the total event rate produced for
$\mathcal{L}=10^{33}cm^{-2}s^{-1}$ at the LHC is $\sim 10^{8}$ events
per second, whereas the event rate for $W$ boson production is $\sim
200$ events per second and for $t \bar{t}$ is $\sim 0.8$ events per
second.  Although there is an improved discovery potential at the LHC
compared to the Tevatron, it will still be a challenge to separate out
the ``interesting'' from the ``uninteresting'' events.

\section{Definitions} \label{sec:defs}
In this section I outline some definitions that all high energy
physicists, both theorists and experimentalists, should know.

\subsection{Rapidity and Pseudorapidity}
The {\it natural} coordinates of a typical collider experiment are
cylindrical around the beam-pipe.  If we assume the $z-$axis to be in
the direction of the beam, we can define $\theta$ as the polar angle
and $\phi$ as the azimuthal angle, and $z=0$ is at the center of the
detector or at the interaction point.  

The rapidity, $y$, of a particle is a function of the energy, E, and
the $z$-component of the momentum, $p_z$ and is defined as:
\begin{eqnarray}
y = \frac{1}{2} \log(\frac{E+p_z}{E-p_z}) = \tanh^{-1}(\frac{p_z}{E}).
\end{eqnarray}

In the coordinate system defined above, the polar angle $\theta$ is not
Lorentz-invariant.  However, what we can define is the pseudorapidity,
$\eta$, as a function of $\theta$ as:

\begin{eqnarray}
\eta \equiv - \log \tan(\theta /2).
\end{eqnarray}

\noindent We can then define the forward region as $\eta \geq 1$ (or
$\theta \approx 0$), the backward region as $\eta \leq -1$ (or $\theta
\approx \pi$) and the central region as $\eta =0$ (or $\theta =
\pi/2$).

Any change in rapidity, $\Delta y$, is Lorentz-invariant
under boosts along the beam direction, and for a massless particle (or
a nearly massless particle where $p>>m$) the rapidity and
pseudorapidity are approximately equal.  It is also interesting to
note, that we can calculate the $\eta$ of a particle without knowing
its mass (which is very handy for experimentalists).

\subsection{$\Delta R$ Distance}

In order to determine the separation in direction between particles, experimentalists use $\Delta R$ as a measure of ``distance'' and is defined as:
\begin{eqnarray}
\Delta R = \sqrt((\Delta \eta)^2 + (\Delta \phi)^2), 
\end{eqnarray}

\noindent where $\Delta \eta$ and $\Delta \phi$ are the particles'
separation in pseudorapidity and azimuthal angle, respectively.  For
example, this is very useful in the reconstruction of ``jets'', where
we use cones of $\Delta R$ to group particles with each other; more on
this in Sec.~\ref{sec:partid}.

\subsection{Transverse Quantities}
Experimentalists also find it useful to focus on quantities measured in
the transverse plane, or the plane perpendicular to the beam $z-$axis.

One quantity that is commonly used is the transverse momentum of a particle, $p_T$, defined as: 
\begin{eqnarray}
p_T = p \sin \theta.
\end{eqnarray}

\noindent Note that the $p_T$ is invariant under $z-$boosts.
Particles that escape detection (or end up in the forward region) have
close to zero $p_T$.  In this sense, the transverse plane is
{\it{opposite}} of forward.

Additional transverse quantities that are often use are the transverse energy, $E_T$:
\begin{eqnarray}
E_T = E \sin \theta,
\end{eqnarray}
\noindent and the transverse mass, $m_T$:
\begin{eqnarray}
{m_T}^2 = \sqrt{{E_T}^2 - {p_T}^2}.
\end{eqnarray}

\noindent  One of the most interesting and most difficult quantities for experimentalists to understand is the missing transverse energy in an event, $\not{\!\!\!E_T}$, defined as:
\begin{eqnarray}
\not{\!\!\!E_T} \equiv - \sum_{i} {E^{\,i}}_{\!T} \hat{n}_i = - \!\!\!\! \sum_{\rm{all ~~ visible}} \! \! \vec{E_T},
\label{eqn:met}
\end{eqnarray}

\noindent where $\hat{n}_i$ is the component in the transverse plane
of a unit vector that points from the interaction point to the
$i^{th}$ calorimeter tower (see Sec.~\ref{sec:cals}).  It is an
event-wide $z$-boost-invariant quantity and many new physics
signatures are expected to show up with large $\not{\!\!\!E_T}$.
Experimentalists also find it interesting to look at the measure of
the scale of the visible $p_T$ in an event, or $H_T$, loosely defined
as:

\begin{eqnarray}
H_T \equiv \sum_{i=objects} | {\vec{p}_{T \, ,i}} |.
\end{eqnarray}

\noindent The definition of $H_T$ varies since it depends on which
objects (leptons, jets, $\not{\!\!\!E_T}$) are included in the sum.
This is also an event-wide $z-$boost-invariant quantity which could
distinguish a SM final state from one produced by new physics.

So why are experimentalists so interested in the transverse plane?
Why not look for missing $p_z$ or missing E?  Unfortunately, in hadron
collisions you do not have the luxury of knowing the initial state
exactly.  Remember what we said in Sec.~\ref{sec:hadroncoll}, the proton itself is
not what scatters.  The particles that do scatter (underlying event)
and escape detection have large $p_z$ so visible $p_z$ is not
conserved and is therefore not a useful variable.  However, to a good
approximation the visible $p_T$ is conserved, which is what makes it
so useful.

\section{Particle Interactions with Matter}\label{sec:partint}

To understand the various LHC detectors (and their differences) first
requires a basic understanding of the interactions of high energy
particles with matter.  Particles can interact with atoms and
molecules, atomic electrons and the nucleus.  These interactions
result in several effects such as ionization, elastic scattering,
energy loss and pair-creation.  There are several respectable sources
on interactions of particles with matter ~\cite{pdg,leo} and the main
one used here is the PDG~\cite{pdg}.

\subsection{Energy Loss of Charged Heavy Particles}

The primary source of energy loss of moderately relativistic heavy
charged particles, such as muons, pions and protons, in matter is via
ionization and atomic excitation.  The average rate of energy loss is
described by the Bethe-Bloch equation:

\begin{eqnarray}
- \frac{dE}{dx} = K z^2 \frac{Z}{A} \frac{1}{\beta^2} [ \frac{1}{2} ln \frac{2 m_e c^2 \beta^2 \gamma^2 T_{max}}{I^2} - \beta^2 - \frac{\delta(\beta \gamma)}{2}],
\label{eqn:betabl}
\end{eqnarray}

\noindent where $z$ is the charge of the particle, $Z$ is the atomic
number of the material the charged particle is traversing, $A$ is the
atomic number of the material, $K= 4 \pi N_A r_e^2 m_e c^2$ ($N_A$ is
Avogadro's number, $r_e$ is the classical electron radius, and $m_e
c^2$ is the mass of the electron), $\beta$ and $\gamma$ describe the
relativistic speed of the particle, $I$ is the mean excitation energy and
$T_{max}$ is the maximum kinetic energy of a free electron in the
collision. Equation~\ref{eqn:betabl} is also referred to as the
{\it stopping power}.  The ionization, $dE / dx$, is typically expressed
in terms of $MeV/(g/cm^2)$ and is dependent on the density of the
material the charge particle is traversing.  The minimum ionization is
found to be at a value of $\beta \gamma \approx 3$, and is independent
of the charged particle's target.

Additionally multiple coulomb scattering off of nuclei is also an important effect
for high energy charged particles since as they ionize while traveling
through materials, they end up changing their direction with each interaction.
The distribution of this multiple scattering is described by a Gaussian 
of width $\theta_0$:
\begin{eqnarray}
\theta_0  = \frac{13.6 MeV}{\beta c p} z \sqrt{x/X_0} [1+ 0.038 ln(x/X_0)],
\label{eqn:moliere}
\end{eqnarray}

\noindent where $\beta c$ is the velocity, $p$ is the momentum, $z$
is the charge of the scattered particle and $x/X_0$ is the thickness
of the material in units of radiation lengths $X_0$ (defined in
Sec.~\ref{sec:emshower}). Equation~\ref{eqn:moliere} holds for
small scattering angles, but for high scattering angles large
non-Gaussian tails appear.

\subsection{Energy Loss of Electrons and Photons and Electromagnetic Showers}
\label{sec:emshower}

Electrons primarily loose energy via bremsstrahlung and ionization.
The rate at which electrons loose their energy by bremsstrahlung is
nearly proportional to its energy and the rate of ionization loss
rises logarithmically.  There is a critical energy, $E_c$, at which the
two loss rates are equal and it depends strongly on the absorber.  For
example, this critical energy for lead is 9.5 MeV.

The characteristic length that describes the energy 
decay of a beam of electrons is called the radiation length, $X_0$, defined as:
\begin{eqnarray}
X_0 = \frac{716.4 g cm^{-2} A}{Z(Z+1) ln(287/\sqrt{Z})},
\end{eqnarray}

\noindent where $A$ is the atomic mass of the material and $Z$ is the
atomic number.  It is the average distance the electron travels until
its energy is reduced by a factor of $1/e$ due to bremsstrahlung.  By
expressing the thickness in terms of $X_0$ the radiation loss is
approximately independent of the material.  The amount of energy loss
of electrons by bremsstrahlung is:
\begin{eqnarray}
- \frac{dE}{dx} = \frac{E}{X_0}.
\end{eqnarray}

\noindent As is shown in Tab.~\ref{tab:radlen}, higher $Z$ materials have
shorter radiation lengths.  For example, lead, which has a density of
$11.4 ~~ g/cm^{3}$ has $X_0 = 5.5$ mm. We will see later that when
designing calorimeters we want as little material as possible in front
of them and high $Z$ materials make good electromagnetic calorimeters.

\begin{table}[t]
\tbl{Radiation lengths and interaction lengths for various materials}
{\begin{tabular}{c c c} \hline
Material & $X_0$ ($g/cm^{2}$) & $\lambda_n$ ($g/cm^{2}$) \\ \hline
$H_2$ & 63 & 52.4\\ 
Al & 24 & 106 \\ 
Fe & 13.8 & 132 \\ 
Pb & 6.3 & 193 \\ \hline
\end{tabular}
}
\label{tab:radlen}
\end{table}

The concept of a radiation length can also be applied to photons.
When high energy photons lose energy in matter they do so via $e^{+}
e^{-}$ pair production.  The mean free path, $\ell$, for pair
production by a photon is:
\begin{eqnarray}
\ell = \frac{9}{7} X_0.
\end{eqnarray}

\noindent For electrons, as we just described, $\ell = X_0$.  So if
we have a high energy photon passing through an absorber, it will
produce electrons, which then radiate bremsstrahlung photons, and so
on, the process repeats.  This electromagnetic cascade of pair
production and bremsstrahlung generate more electrons and photons with
lower energy and is referred to as an electromagnetic shower. The
transverse (lateral) development of electromagnetic showers scale with
what is referred to as the Moli\`{e}re radius, $R_M$: ~\cite{pdg}

\begin{eqnarray}
R_M = 21 MeV X_0 / E_c, 
\end{eqnarray}

\noindent where $E_c$ is the critical energy as described above.

\subsection{Hadronic Showers}

Hadronic showers are produce by interactions of heavy particles with
nuclei.  These showers are described by the nuclear interaction
length, $\lambda_n$:

\begin{eqnarray}
\lambda_n \approx 35 g cm^{-2} A^{1/3}.
\end{eqnarray}

\noindent For heavy, or high Z, materials the nuclear interaction length is quite a
bit longer than the electromagnetic one and $\lambda_n > X_0$ (see Tab.~\ref{tab:radlen}).  This
results in hadronic showers starting later than electromagnetic
showers and are more diffuse.  For example, from
Tab.~\ref{tab:radlen} lead, which has a density of =11.4 $g/cm^3$, has
an interaction length of $\sim$ 17 cm.

\section{Particle Detectors}\label{sec:partdet}

The goal of every collider experiment is to completely surround
the collision by arranging layers of different types of subdetectors.  In
Sec.~\ref{sec:partint} we just learned how different particles
interact with matter so in order to identify them we exploit these
differences.  The key information of the particles that we want to
extract is their momentum and charge, their energy, and their species.

\begin{figure}[t]
\begin{center}
\psfig{file=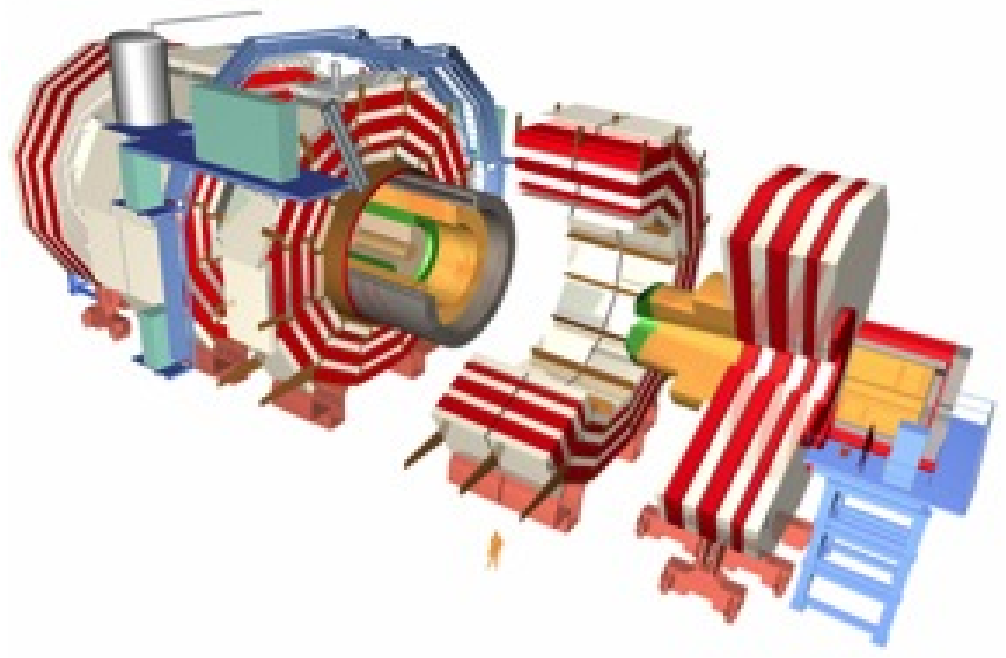,width=3.in}
\psfig{file=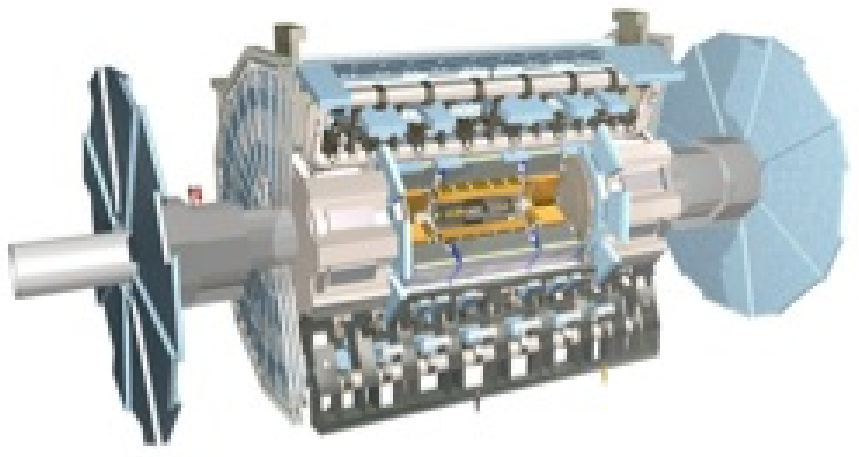,width=3.in}
\caption{Schematic drawings of the CMS (top) and ATLAS (bottom) detectors.}
\label{figs:cmsatlas}
\end{center}
\end{figure}

Figure~\ref{figs:cmsatlas} shows schematic drawings of the CMS (top)
and ATLAS (bottom) detectors, which have the traditional layered
detector structure.  These detectors have the following general features, starting from center moving outwards:
\begin{itemize}
\item {\it Tracking detectors within a magnetic field:} measures the charge, trajectory and momentum of charged particles 
\item {\it Electromagnetic calorimeter:} measures the energy and position of electromagnetic particles
\item {\it Hadronic calorimeter:} measures the energy and position of hadronic particles
\item {\it Muon chambers:} measures the trajectory and momentum (along with the tracking detector) of muons 
\end{itemize}

\noindent In the sections below I provide a brief description of these
detectors, giving examples from both the LHC experiments and Tevatron
experiments.  Additional details on particle physics detectors can be
found in these Refs.~\cite{pdg, leo}~.  I summarize the detector
technologies used in the CMS and ATLAS detectors in
Sec.~\ref{sec:cmsatlas}.

\subsection{Tracking Detectors}
\label{sec:tracking}

\begin{figure}[t]
\begin{center}
\psfig{file=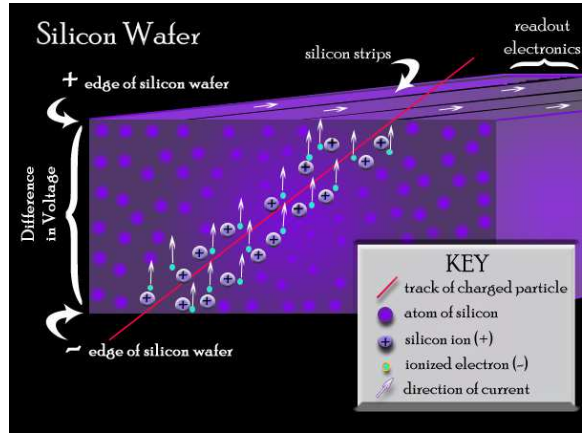,width=3.in}
\caption{Schematic of a silicon semiconductor detector \cite{silicon}.}
\label{figs:sili}
\end{center}
\end{figure}

\begin{figure}[t]
\begin{center}
\psfig{file=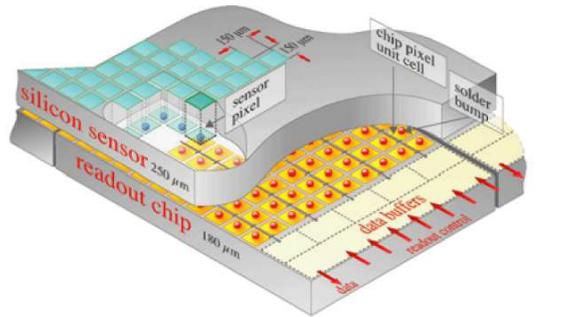,width=3.in}
\caption{Schematic of the silicon pixel detector at CMS \cite{cmspix}.}
\label{figs:pix}
\end{center}
\end{figure}

The main goal of tracking detectors is to measure the momentum, charge
and trajectory of charged particles.   Ideally, we we want tracking
detectors to contain as little material as possible in order to
minimize multiple scattering.  There are two main technologies of
tracking detectors in particle physics:

\begin{itemize}

\item {\it Gas/wire drift chambers:} These devices are made of wires
  in a volume filled with a gas, such as Argon-Ethane.  They measure
  where a charged particle has crossed when it ionizes the gas.  There
  is an electrical potential applied to the wires so atomic electrons
  knocked off the atoms in the gas drift to a positively charged sense
  wire.  The chamber are connected to electronics which measure the
  charge of the signal and when it appears.  To reconstruct the tracks
  of the charged particles several chamber planes are necessary.
  Advantages to drift chambers is their low thickness (in terms of
  $X_0$) and are the traditionally preferred technology for large
  volume detectors.  Typical single hit resolutions range from $\sim
  100-200 \mu m$.  An example of such a device is the CDF experiment's
  Central Outer Tracker (COT) ~\cite{cdfcot} which has approximately
  30,000 wires.

\item {\it Silicon detectors:} Silicon detectors are semi-conductor
  detectors which are modified by doping.  For example, doping with
  Antimony gives an n-type semiconductor or with Boron which gives a
  p-type semiconductor.  This doped silicon is then used to create a
  p-n junction, to which a very large reverse-bias voltage is applied.
  This creates a ``depletion zone'' and once the silicon device is
  “fully depleted” we are left with an electric field.  When charged
  particles cross the detector they ionize the depletion zone and
  create an electrical signal.  Figure~\ref{figs:sili} shows a
  schematic drawing of a charged particle interacting in a silicon
  device, which has a typical thickness of $\sim 300 \mu m$.  Silicon
  detectors come in two varieties, either metal strips (as seen in
  Fig.~\ref{figs:sili}) or pixels (shown in Fig.~\ref{figs:pix}) which
  provide much higher granularity and a higher precision set of
  measurements.  For example, the CMS silicon strip resolution ranges
  from $8-64 \mu m$ and for its pixel detector is $\sim 15-30 \mu m$.
  Additionally, the number of pixel sensor channels at CMS is $\sim
  65$ million and at ATLAS is $\sim 80$ million. These detectors are
  radiation hard and are important for detection secondary vertices
  (for example, from b-hadron decays as described in
  Sec.~\ref{sec:partid}) close to the primary interaction.  Silicon is
  now the dominant sensor material in use for tracking detectors at
  the LHC and especially for CMS.
\end{itemize}

Since a magnetic field is applied within the detector the momentum and
charge of the particle is measured using a few points of the
particle's track (trajectory) which we can use reconstruct the
curvature of the track.  The transverse momentum ($p_T$) of charged
particles is proportional to the radius of curvature and to the $B$
field.  In particular,

\begin{eqnarray}
p_T = 0.3 ~q ~B ~r,
\end{eqnarray}
 
\noindent where the reconstructed track $p_T$ is measured in GeV/c, $B$
is in Tesla, the total particle charge is $q e$ (e is the magnitude of
the electron charge) and $r$ is measured in meters and is the radius of curvature
of the track.

\subsection{Electromagnetic and Hadronic Calorimeters}
\label{sec:cals}

Electromagnetic calorimeters are designed to measure the energy of
electromagnetic particles (both charged and neutral) and their
position.  This is done by constructing them using a heavy, high $Z$
material to initiate an electromagnetic shower, as described in
Sec.~\ref{sec:emshower}, to totally absorb the energy and stop the
particles.  The important parameter for the material used in
electromagnetic calorimeters is the radiation length $X_0$, and have
typical values of 15-30 $X_0$.  Additionally, it is key to have as
little material before the calorimeter as possible (this means the
tracker) so that the particles do not radiate before they reach it.

The relative energy uncertainty (or resolution), $\sigma_E$, of
calorimeters decreases with the energy $E$ of the particle and can be
parameterized as follows:

\begin{eqnarray}
\frac{\sigma_E}{E} = \frac{a}{\sqrt{E}} \oplus b \oplus \frac{c}{E},
\label{eqn:enres}
\end{eqnarray}

\noindent where $a$ is referred to as the stochastic term and
quantifies statistics related fluctuations, $b$ is the constant term
and $c$ is primarily due to noise (for example, in the electronics).
The three terms in Eq.~\ref{eqn:enres} are added in quadrature
(denoted by the symbol $\oplus$).

\begin{figure}[t]
\begin{center}
\psfig{file=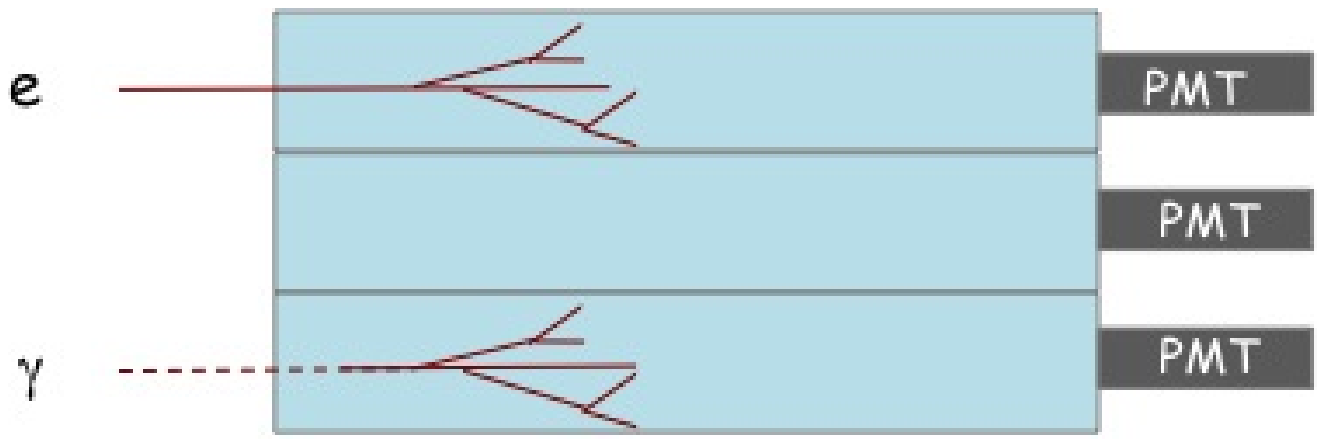,width=3.in}
\psfig{file=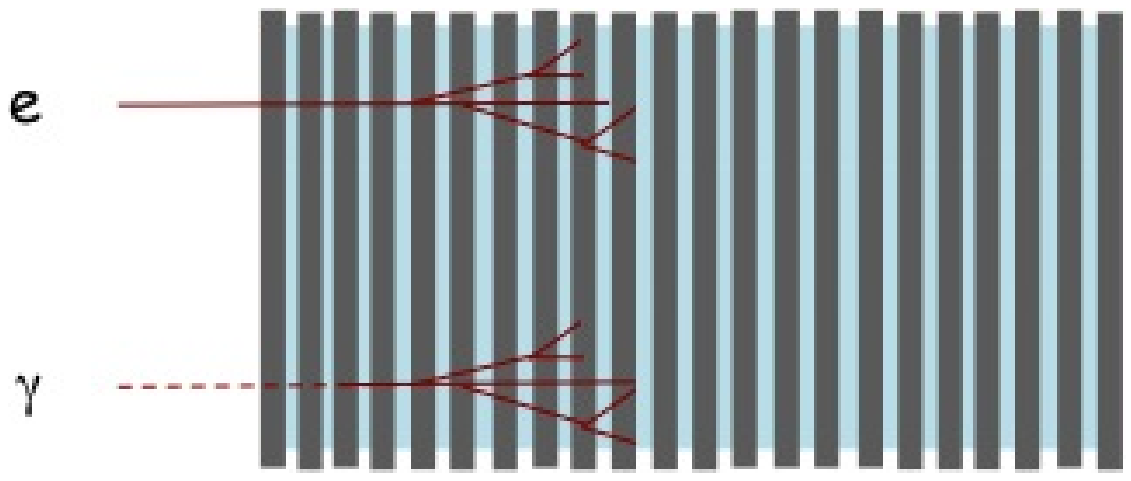,width=3.in}
\caption{Schematics of a homogeneous calorimeter (top) and a sampling calorimeter (bottom).}
\label{figs:cal}
\end{center}
\end{figure}

There are two types of calorimeter detectors:
\begin{itemize}
\item {\it Homogeneous calorimeter:} These detectors are generally
  made of an inorganic heavy, high $Z$ material which is also
  scintillating.  The idea is to create an entire volume to generate
  the electromagnetic signal, as seen in Fig.~\ref{figs:cal} (top).
  Examples of these calorimeters include a variety of crystals such as
  CsI, NaI, and PbWO, and ionizing noble liquids such as liquid Ar.
  Energy resolutions of these types of detectors are typically
  $\frac{\sigma_E}{E} \sim 1\%$.

\item {\it Sampling calorimeter:} These calorimeters are made of an
  active medium which generates signal and a passive medium which
  functions as an absorber as seen in Fig.~\ref{figs:cal} (bottom).
  Examples of active medium materials are scintillators, ionizing
  noble liquids, and a Cherenkov radiator.  The passive material is
  one of high density, such as lead, iron, copper, or depleted
  uranium.  Energy resolutions of sampling calorimeter detectors are
  typically $\frac{\sigma_E}{E} \sim 10\%$.
\end{itemize}

\noindent The scintillating light created in calorimeters is
interpreted as a signal using photo-multiplier tubes (PMT's) and translated as
the energy of the particle.

The purpose of hadronic calorimeters is to measure the energy of heavy
hadronic particles. Hadronic calorimeters are similar to
electromagnetic calorimeters but in this case the important parameter
of the absorber is the interaction length $\lambda_n$.  In general, a
hadronic calorimeter has $\lambda_n \approx 5-8$.  They typically are
sampling calorimeters and tend to be larger and coarser in sampling
depth than electromagnetic calorimeters, and therefore have larger
energy resolutions.  For example, the stochastic term is usually in
the 30-50\% or even higher.

\subsection{Muon Chambers}

Recall that the muon signature is that of a minimum ionizing particle
and extraordinarily penetrating and therefore the detectors for
identifying them are the outer-most layer of a collider detector.
These detectors are made up of several layers of tracking chambers as
described in Sec.~\ref{sec:tracking}.  Their primary purpose is to
measure the momentum and charge of muons.  The measurements from the muon
chambers are combined with the tracks reconstructed with the inner
tracker to fully reconstruct the muon trajectory.

Muon chambers in LHC experiments are made from a series of different
types of tracking chambers for precise measurements and some examples include:
\begin{itemize}
\item Drift Tubes (DT's): Wire chamber devices, so when muons
  traveling through kick off atomic electrons in the gas and drift to
  the positively charged wire.
\item Cathode Strip Chambers (CSC's): Wires crossed with metallic
  strips in a gas volume, so when muons traverse the detectors
  electrons drift to the positively charged wire as described above.
  Additionally, the positive ions in the gas drift to the metallic
  strips and induce a charged pulse perpendicular to the wire, giving
  a two dimensional coordinate of the traveling muon.
\item Resistive Plate Chambers (RPC's): Oppositely charged parallel
  plates containing a gas volume, creating drift electrons when muons
  cross the detector.
\end{itemize}

\begin{figure}[t]
\begin{center}
\psfig{file=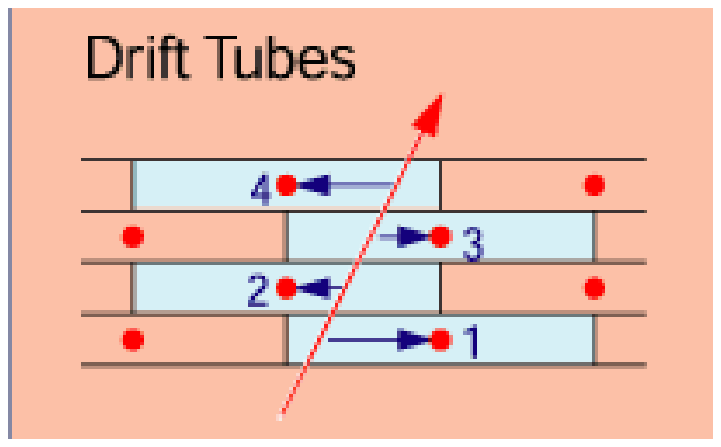,width=1.5in}
\psfig{file=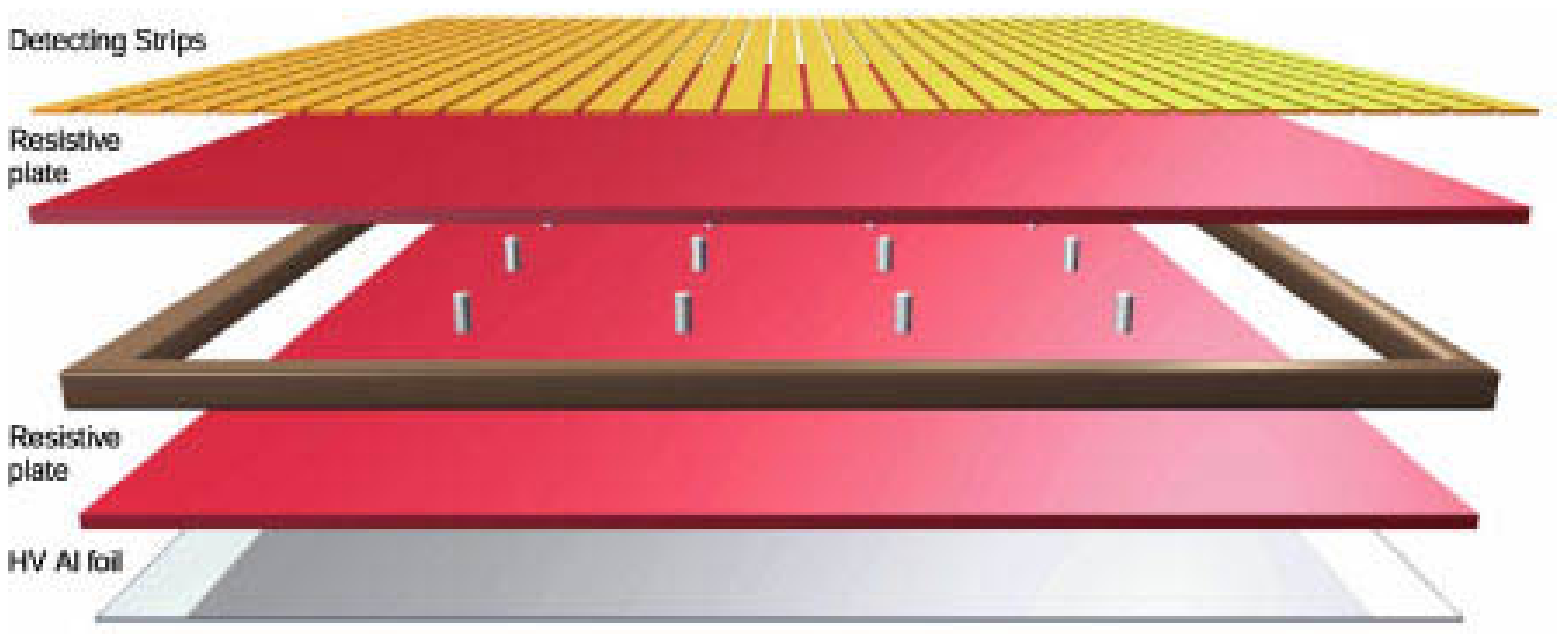,width=1.5in}
\psfig{file=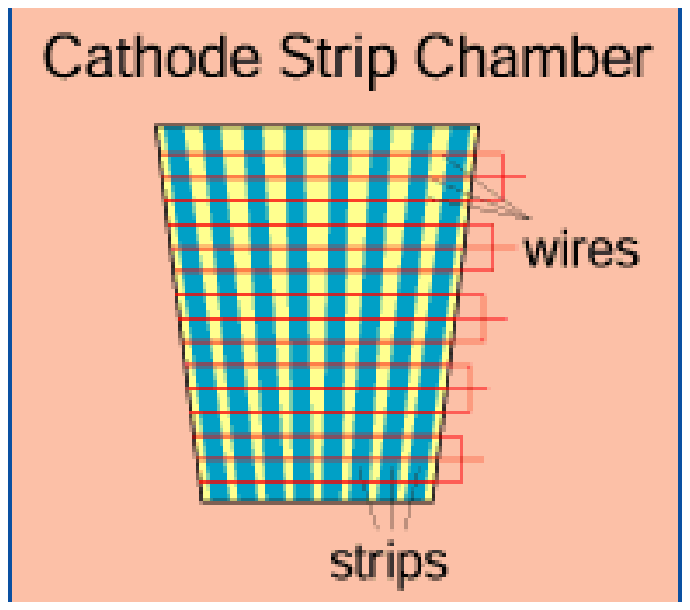,width=1.in}
\caption{Examples of Muon Chambers in the LHC experiments \cite{cmsmudet}.  From left to right:  Drift Tubes, Resistive Plate Chambers and Cathode Strip Chambers.}
\label{figs:muondet}
\end{center}
\end{figure}

\noindent Schematic drawings of DT's, RPC's and CSC's are shown in
Fig.~\ref{figs:muondet}. 

\subsection{The ATLAS and CMS Detectors}
\label{sec:cmsatlas}

In this section I give a brief summary of the details of the CMS and
ATLAS detectors shown in Fig.~\ref{figs:cmsatlas}.  Additional
detailed information can be found in the technical design reports
(TDR) ~\cite{cmstdr, cmstdrurl,atlastdr, atlastdrurl} for the two
experiments.

Both the CMS and ATLAS detectors are large scale experiments in every
sense.  CMS is 21 m long, 15 high m and 15 m wide and weighs 12,500
tons.  The dimensions of ATLAS are even larger, 46 m long, 25 m high
and 25 m wide, and weighs 7000 tons.  CMS is located at Point 5 around
the LHC ring in Cessy, France, whereas ATLAS is located at Point 1 and
is in Meyrin, Switzerland (see Fig.~\ref{figs:lhcacc}).  Due to the
high intensity of the collisions the detectors will experience, they
both have been designed to be very radiation hard, in particular the
tracking detectors closest to the beam-pipe.  The ATLAS and CMS
experiments have designed their subdetectors using different
approaches and a summary of the detector technologies used is shown in
Tab.~\ref{tab:lhcdet}.

\begin{table}[t]
\tbl{ATLAS and CMS Subdetectors}
{\begin{tabular}{c c c} \hline
Detector & ATLAS & CMS \\ \hline
Tracking & silicon/gas & silicon \\ 
EM Cal & liquid Argon & PbWO \\
HAD Cal & steel/scintillator & brass/scintillator \\
Muon & RPC's/drift & RPC's/drift \\
Magnet & Solenoid (inner) / Toroid (outer) & Solenoid \\
B Field & $\sim$ 2 T / 4 T & $\sim$ 2 T \\ \hline
\end{tabular}
}
\label{tab:lhcdet}
\end{table}

\section{Particle Identification}
\label{sec:partid}
\begin{figure}[t]
\begin{center}
\psfig{file=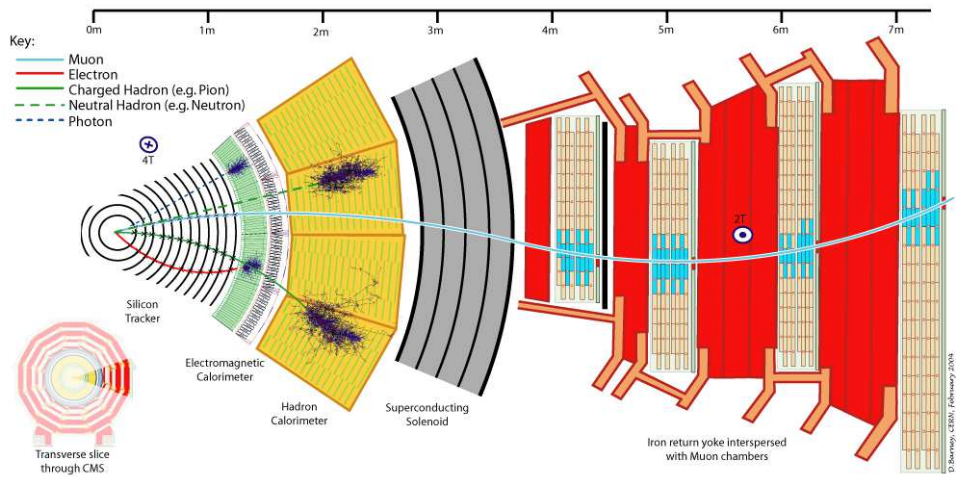,width=5in}
\caption{Transverse slice of the CMS detector, showing the identification of particles~\cite{cmsurl}.}
\label{figs:cmspart}
\end{center}
\end{figure}

In this section, I describe how these detectors described in
Sec.~\ref{sec:partdet} are used for particle identification.
Figure~\ref{figs:cmspart} shows a schematic of a transverse slice of
the CMS detector outlining the identification of various particles.
One may find it useful to refer to this diagram while reading the
description below.

\begin{itemize}

\item {\it Electrons and Photons:} Electrons are identified as an
  energy deposit in the electromagnetic calorimeter, and is required
  to have a shower shape (energy loss) consistent with an electromagnetic
  shower.  It is also required to have little or no energy in the
  hadronic calorimeter.  Since electrons are charged particles it
  needs to be associated with a track reconstructed in the tracker,
  and is therefore required to have a matched position measurement in
  the calorimeter with the one from the track.  If the electromagnetic
  cluster of energy does not have a track pointing to it then it
  becomes a candidate for being a photon.

\item {\it Muons:} Muon identification begins by reconstructing a track
  in the muon system which is then matched with a track in the inner
  tracker.  Additionally, since muons are minimum ionizing particles,
  they are expected to deposit little or no energy in the
  calorimeters.

\item {\it Jets:} Jets are created when a quark or gluon is knocked
  out of the proton and due to parton confinement subsequently a
  hadron is created.  This hadron forms a jet once it decays and
  fragments into many particles (hadronization), which are essentially
  collimated object.  The reconstruction of a jet is the
  experimentalist’s representation of a parton.  There are several
  algorithms for reconstructing jets but overall what these
  reconstruction algorithms do is attempt to group the particles from
  the hadronization process together and measure the energy of the
  parton.  There are two main categories of jet algorithms that
  experimentalists and theorist use to reconstruct jets: (1) Cone
  algorithms~\cite{cone} when one draws circles of $\Delta R$ around
  clusters of energy according to some rule, and (2) Recursive
  cluster reconstruction such as the anti-$k_{T}$ algorithm
  ~\cite{antikt} which is now the default jet algorithm of choice for
  the LHC experiments.  

  Measuring the jet energy has several challenges since it is
  impossible to determine which particles came from which
  hadronization process.  There are several effects which contribute
  to the complication of the jet energy measurement, such as multiple
  $p p$ interactions, spectator partons interacting and noise in the
  calorimeters.  However, experimentalists have ways of correcting for
  such effects and this calibration the jet energy is generally called
  the “Jet Energy Scale” (JES) and often depends on the $p_T$ and the
  $\eta$ of the jet.

\item {\it b-hadrons:} There is a special category of jets coming
  from $b$ hadrons which are long-lived (with $c\tau \sim 450 \mu m$)
  and massive.  There are two standard techniques for identifying a $b$
  hadron decay, referred to as {\it b-tagging}.  One can look for
  displaced tracks forming a secondary vertex away from the primary
  vertex of the interaction.  Alternatively one can identify soft
  leptons (electrons or muons) inside the jet, which would be a
  signature specific to semi-leptonic $b$ decays.

\item {\it Tau Leptons:} The identification of tau leptons is for
  hadronically decaying taus, which decay $\sim 49\%$ of the time to a
  single charged hadron and neutrinos and $\sim 15\%$ of the time to
  three charged hadrons and neutrinos.  Leptonically decaying taus are
  indistinguishable from ``normal'' electrons and muons.  The
  reconstruction algorithms for taus assume that taus form narrow jets
  in the calorimeter.  First one forms a $\Delta R$ cone around
  clusters of energy and tracks (a signal cone) and a second larger
  $\Delta R$ cone around the signal cone (an isolation cone) where
  there is little or no calorimeter and track activity.  In the signal
  cone, one or three tracks are required as well as electromagnetic
  energy in the calorimeters from neutral particles (such as
  $\pi^0$s).

\item {\it Neutrinos or $\not{\!\!\!E_T}$:} Neutrinos are weakly
  interacting particles and pass through all the material in the LHC
  detectors.  They are identified indirectly by the imbalance of
  energy in calorimeters.  This missing energy was previously defined
  above in Eq.~\ref{eqn:met} and recall that it is one of the
  most interesting and most difficult quantities for experimentalists.
  Various effects could contribute to the complications of the
  $\not{\!\!\!E_T}$ measurement such as dead calorimeter cells or a
  jet whose hardest hadron enters a crack (between cells) in the
  calorimeter or an improperly calibrated calorimeter.  Therefore, we
  need to carefully understand this quantity as it is very important
  for searches of new physics processes which could produce additional
  weakly interacting particles.

\end{itemize}

\section{Selecting and Storing the Interesting Events: Trigger and Computing} \label{sec:trig}

At design a center-of-mass energy of $14$ TeV and a luminosity of $10^{33} cm^{-2}
s^{-1}$, Fig.~\ref{figs:smxsec} shows that the total cross section at the LHC
will be $\sim 10^8$ nb. The rate for all collisions will be around 40
MHz.  Since it is not possible to record every collision event, quick
decisions need to be made {\it a priori} selecting the interesting events
worthy of analysis.  This filter, or trigger, needs to single out rare
processes and reduce the common processes.  We also want to keep
“less” interesting events for ``standard-candle'' measurements (such
as jet and $W$ boson and $Z$ boson production cross sections),
calibrations, and so on. It is critical to consider carefully the
make-up of the trigger and make wise choices, otherwise the events
will be thrown away forever.

A {\it typical} trigger table will contain triggers on: electroweak
particles (photons, electrons, muons, taus) at as low an energy as
possible, very high-energy partons (jets), and apparent invisible
particles ($\not{\!\!\!E_T}$).  Theory very often plays a role in
guiding these choices, therefore it is important to have good
communication between theorists and experimentalists.

The LHC experiments have two levels of triggers, one which bases its
decision on hardware electronics (L1), and a second level which based
on software programming (the high level trigger or HLT).  Recall, the
starting trigger rate is 40 MHz, which gets reduces after the L1
trigger to a rate of around 100kHz.  The HLT trigger further prunes
this down to roughly 150-200 Hz, which is the event rate that the
experiments record.  Therefore, the final decision of the trigger is
to keep $\sim$ 1/200,000 events occur every second, there is no room
for mistakes.

One should be very aware that all measurements are distorted by the
trigger selection thresholds and any measurement must account for the
efficiency of the trigger and that resulting distortion.  Therefore,
it is necessary to include ``backup'' or ``monitoring'' trigger for
measuring the efficiencies of the more interesting triggers to be used
for physics analysis.  Additional details on the ATLAS and CMS trigger
and data acquisition systems can be found in Refs. ~\cite{cmstdr,
  cmstdrurl,atlastdr, atlastdrurl}.

The LHC will produce roughly 15 petabytes (15 million gigabytes) of
data annually.  Finally, there is the challenging task of distributing
the recorded data around the world for analysis.  The LHC has a tiered
computing model to distribute this data around the world, referred to
as the Grid~\cite{gridurl}.

\section{Status of the LHC}
\label{sec:lhcstat}
These TASI lectures were given in June 2009 and the LHC has since turned
on and the experiments have been collecting data.  It was on November
20, 2009 when the LHC first came back online, circulating proton beams
of 450 GeV and three days later the beams collided for the first time
at at $\sqrt{s}=900$ GeV.  Then in December, protons collided for the
first time ever at $\sqrt{s}=2.36$ TeV, exceeding the center-of-mass
energy of the Tevatron.  And on March 30, 2010 the LHC achieved again
the highest ever energy collisions at $\sqrt{s}=7$ TeV, and a new era in
particle physics commenced.  Since then the machine and the
experiments have been running smoothly and has so far achieved a
luminosity around $10^{27} cm^{-2}s^{-1}$, with a goal of $\sim
10^{32} cm^{-2}s^{-1}$.  The plan is to continue running the
accelerator at $\sqrt{s}=7$ TeV through 2011 (with a short technical
stop at the end of 2010) until it has delivered $> 1 fb^{-1}$ of good
collision data to the experiments.  This dataset will be enough to
make potentially very exciting new discoveries in the near
future.

\section{Early LHC Physics Measurements} \label{sec:lhcphys}
With the initial data from the LHC, in order to have confidence of any
potential claims of discovery, the very first job of the
experimentalists is to understand the detector.  The early LHC
measurements will be focused on calibrating, and aligning the detector
as well as rediscovering the SM since it is the SM particles which are
the only ones we are {\it guaranteed} to see.  A complete list of
expectations for physics measurements from the CMS and ATLAS
experiments can be found in their TDR's ~\cite{cmsphystdr,
  cmstdrurl,atlastdr, atlastdrurl} and updated results located at the
experiment websites~\cite{cmsurl,atlasurl}.

Here I will only list a few examples of early LHC physics
measurements of SM processes.  Without measurements such as the ones
listed below, we can not be assured of any claims of discovery of new
physics.
\begin{itemize}
\item Charged track track multiplicity: This measurement has already
been made by the ALICE~\cite{alicecpm}, CMS~\cite{cmscpm} and
ATLAS~\cite{atlascpm} experiments with the recently collected 900 GeV
and 2.36 TeV data.
\item Inclusive jet cross section 
\item $Z$ and $W$ boson production cross sections 
\item $t \bar{t}$ pair production cross section
\end{itemize}

In the next section, I will highlight a few of the key elements which
go in making an example measurement such as a production cross
section.

\subsection{Example Analysis: Measuring a Cross Section}
\label{sec:xsec}
Measurements of the production cross sections of known processes
produced in high energy $pp$ collisions provide important tests of the
SM.  Although the measurement of a cross section is primarily a
“counting experiment”, one should not be fooled into thinking it is a
simple analysis; it is actually rather complex with many ingredients
which need detailed understanding.  Experimentally, the cross section
for a process of interest is measured as:

\begin{eqnarray}
\sigma = \frac{N_{obs} - N_{bkg}}{ \int \mathcal{L} dt \times \epsilon },
\end{eqnarray}

\noindent where $N_{obs}$ is the number of observed candidate events
selected in the data sample, $N_{bkg}$ is the estimated number of
background events mimicking the signal, $\int \mathcal{L} dt$ is the
integrated luminosity of the data sample analyzed, and $\epsilon$ is
the overall efficiency of observing the produced events of interest.

The evaluation of the background processes which {\it fake} your
signal can often be a difficult task.  In general, backgrounds are
evaluated using a combination of Monte Carlo simulations of select
processes (such as electroweak production) and evaluating them
directly from the data (for example, jets {\it faking} leptons).

The total efficiency, $\epsilon$, typically has several components.  Overall, one needs to evaluate:
\begin{eqnarray}
\epsilon = \frac{{\rm Number ~of ~events ~used ~in ~the ~analysis}}{{\rm Number ~of ~events ~produced}}.
\end{eqnarray}
\noindent Some of the key ingredients to evaluating the total efficiency are the product of:
\begin{itemize}
\item {\it Trigger efficiency:} Modeling of the trigger in collider
  experiments has been found to be quite complex.  Generally, trigger
  efficiencies are obtained from data, measuring the efficiencies of
  the different components that make up a particular trigger from an
  other trigger with looser requirements.  These efficiencies can have
  a dependence on $p_T$ or $\eta$ for example and a trigger turn-on
  curve as a function of these variables need to be evaluated.
\item {\it Particle identification efficiency:} The identification of
  the final state particles as described in Sec.~\ref{sec:partid}
  are often not highly efficient.  One needs to determine how often an
  object that should have been identified failed the selection
  criteria.  The identification efficiencies are generally obtained
  from data.  For example, for leptons, $Z \rightarrow \ell \ell$ decays
  are ideal for measuring their efficiencies.  $Z$ boson decays provide a
  clean environment and a precisely known mass resonance.  The efficiencies
  are then measured by selecting $Z$ candidate events where only one
  of the leptons is rigorously identified, while the other lepton has
  its selection criteria significantly loosened, and then counting how
  often the loose lepton fails the full selection.
\item {\it Reconstruction efficiencies:} Again, experimentalists rely
  on the data to help them evaluate the efficiency of the
  reconstruction of tracks, the reconstruction of clusters in the
  calorimeters, etc.
\item {\it Kinematic acceptance:} An additional ingredient to knowing
  the total efficiency $\epsilon$ is also knowing the fraction of the
  decays which satisfy the geometric constraints of the detector (for
  example $\eta$ coverage) and the kinematic constraints (for example
  $E_T$ or $p_T$ of the final state objects) of the event selection
  criteria.  The acceptance is primarily determined from a Monte Carlo
  simulation of the signal process.
\end{itemize}

There are a lot of examples of cross section measurements in the
available literature which describe in detail the complexities of the
analysis, and I provide a reference to rather complete one of the
inclusive $W$ and $Z$ cross sections at the Tevatron ~\cite{wzxsec}
for further reading.

\section{Concluding Remarks}\label{sec:conc}

With the startup of the LHC, we are at the dawn of a new era of
particle physics.  In these lectures, I was only able to touch the
surface of the challenges experimentalists face when trying to
understanding the data to the point of confidently making a discovery.
With these lectures I provide a starting point for understanding the
physics of how particles interact with matter ~\cite{pdg, leo} and how we
exploit those interactions in the state-of-the-art CMS and ATLAS
detectors ~\cite{cmstdr, cmstdrurl,atlastdr, atlastdrurl} to be used
in analyses.  It is a great time to be working on the energy frontier
as we are all looking forward to upcoming discoveries at the LHC.

\section{Acknowledgments}\label{sec:ackn}
I would like to thank the TASI organizers for their hospitality and
for their kind invitation to give these lectures.  It was a wonderful
opportunity and I hope I interact again in the near future the highly
enthusiastic group of students who attended the lectures.  Their
energy was refreshing and I encourage them to continue to have
lively discussions with experimental colleagues.

\bibliographystyle{ws-procs9x6}
\bibliography{tasi09_bib}

\begin{thebibliography}{10}

\bibitem{sm}
\url{http://www.symmetrymagazine.org/images/200502/standard3.gif}.

\bibitem{tasi09}
https://physicslearning2.colorado.edu/tasi/tasi\_2009/tasi\_2009.htm.

\bibitem{lhc}
\url{http://cdsweb.cern.ch/}.

\bibitem{cdf}
\url{http://www-cdf.fnal.gov/}.

\bibitem{kidonakis}
N. Kidonakis, ``Higher-order corrections to top-antitop pair and single top
  quark production'', arXiv:hep-ph/0909.0037.

\bibitem{pdf}
J.M. Campbell, J.W. Huston, W.J. Stirling, Rept. Prog. Phys. {\bf 70}, 89
  (2007). Also, see arXiv:hep-ph/0611148.

\bibitem{cteq6m}
J. Pumplin, D.R. Stump, J. Huston, H.L. Lai, S. Kuhlmann, J.F. Owens and W.K.
  Tung, ~J. High Energy Phys. {\bf 0310}, 046 (2003).

\bibitem{mrst2006}
A.D. Martin, W.J. Stirling, R.S. Thorne and G. Watt, Phys. Lett. B {\bf 652}
  292 (2007).

\bibitem{mstw}
A.D. Martin, W.J. Stirling, R.S. Thorne, G. Watt,``Parton distributions for the
  LHC'', arXiv:hep-ph/0901.0002.

\bibitem{pdfmaker}
\url{http://durpdg.dur.ac.uk/hepdata/pdf3.html}.

\bibitem{pdg}
C. Amsler {\it et al.} (Particle Data Group), ``The Review of Particle
  Physics'', Physics Letters B667, 1 (2008) and 2009 partial update for the
  2010 edition.

\bibitem{leo}
W.~R. Leo, {\em Techniques for Nuclear and Particle Physics Experiments}, 2nd
  edn. (Springer, 1994).

\bibitem{cdfcot}
T. Affolder {\it et al.}, Nucl. Instrum. Meth. {\bf A526} 249 (2004).

\bibitem{silicon}
http://www-cdf.fnal.gov/virtualtour/silicon$\_$detector.html.

\bibitem{cmspix}
http://cms.web.cern.ch/cms/Detector/Tracker/index.html.

\bibitem{cmsmudet}
http://cms.web.cern.ch/cms/Detector/Muons/index.html.

\bibitem{cmstdr}
``CMS Physics TDR: Volume I, Detector Performance and Software'',
  CERN-LHCC-2006-001.

\bibitem{cmstdrurl}
http://cmsdoc.cern.ch/cms/cpt/tdr/.

\bibitem{atlastdr}
``ATLAS TDR: Detector and Physics Performance Technical Design Report'',
  CERN-LHCC-1999-14/15.

\bibitem{atlastdrurl}
http://atlas.web.cern.ch/Atlas/internal/tdr.html.

\bibitem{cmsurl}
http://cms.cern.ch/.

\bibitem{cone}
G.C. Blazey {\it et al.}, ``Run II jet physics'', arXiv:hep-ex/0005012.

\bibitem{antikt}
M. Cacciari, G. P. Salam, G. Soyez, J. High Energy Phys. {\bf 04}, 063 (2008).

\bibitem{gridurl}
http://lcg.web.cern.ch/lcg/.

\bibitem{cmsphystdr}
``CMS Physics TDR: Volume II, Physics Performance'', CERN-LHCC-2006-021, J.
  Phys. G, {\bf 34}. 995 (2007).

\bibitem{atlasurl}
http://atlas.ch/.

\bibitem{alicecpm}
The ALICE Collaboration, Eur.Phys.J. {\bf C65}, 111 (2010).

\bibitem{cmscpm}
The CMS Collaboration, J. High Energy Phys. {\bf 02}, 041 (2010).

\bibitem{atlascpm}
The ATLAS Collaboration, arXiv:hep-ex/1003.3124.

\bibitem{wzxsec}
A. Abulencia {\it et al.}, J. Phys. G: Nucl. Part. Phys. {\bf 34}, 2457 (2007).

\end{thebibliography}

\end{document}